\documentclass[pre,showpacs,preprint]{revtex4}

\usepackage{amsmath}

\usepackage{graphicx}
\usepackage{dcolumn}
\usepackage{bm}

\begin{document}

\title{The homogeneous steady state of a confined granular gas }
\author{J. Javier Brey, M.I. Garc\'{\i}a de Soria, P. Maynar,  and V. Buz\'{o}n}
\affiliation{F\'{\i}sica Te\'{o}rica, Universidad de Sevilla,
Apartado de Correos 1065, E-41080, Sevilla, Spain}
\date{\today }

\begin{abstract}
The non-equilibrium statistical mechanics and kinetic theory for a model of a confined quasi-two-dimensional gas of inelastic hard spheres is presented. The dynamics of the particles includes an effective mechanism to transfer the energy injected in the vertical direction to the horizontal degrees of freedom. The Enskog approximation is formulated and used as the basis to investigate the temperature and the distribution function of the steady state eventually reached by the system. An exact scaling of the distribution function of the system having implications on the form of its moments  is pointed out. The theoretical predictions are compared with numerical results obtained by a particle simulation method,  and a good agreement is found.

\end{abstract}


\maketitle

\section{Introduction}
\label{s1}
Granular gases \cite{Ca90,Ha83} are often modeled as an assembly of inelastic hard spheres or disks. In the simplest versions, the energy lost in each collision  is  a fixed fraction of the part of the kinetic energy associated with the normal components of the velocity of the particles at the collision. In other words, the inelasticity in collision is characterized by a constant, velocity independent, coefficient of normal restitution. By extending the methods of kinetic theory and non-equilibrium statistical mechanics, hydrodynamic equations for this model  have been derived for  single component systems in the low-density limit \cite{BDKyS98,ByC00} and for dense fluids \cite{JyR85,GyD99,DByB08,BDyB08}. Also binary gases have been considered \cite{JyM87,GyD02}. An extension of the formalism, valid for the possible scattering laws for dissipative collisions consistent with conservation of momentum and angular momentum, has been worked out \cite{Lu04,Lu05}. This includes as  particular cases, models with a velocity-dependent  coefficient of restitution, which are more realistic \cite {KyK87,RPByS99,ByP04}. Here, the simplest version with a constant coefficient of normal restitution will be considered.

Due to the loss of energy in collisions, there is no equilibrium state for a granular gas. Instead, there is a reference homogeneous state whose temperature decreases monotonically in time \cite{Ha83}. A steady state can only exist if energy is continuously supplied to the system. In real experiments, this can be done by injecting energy through the boundaries, e.g. vibrating some of them \cite{BRyM01}, or by means of an external field, e.g. a granular flow down an  inclined plane \cite{SEGHLyP01}. The price to be paid in both cases is that the steady states are highly inhomogeneous. A theoretical alternative is to introduce some external noise force acting on each particle, i.e. a stochastic thermostat \cite{vNyE98,PLyM99,PTvNyE02}. Nevertheless, the relationship of this mechanism and any experimental set up is rather unclear.

An interesting two-dimensional granular model leading to the existence of an homogeneous steady state has been recently proposed by Brito {\em et al.} \cite{BRyS13}. The model is designed to describe the horizontal dynamics of a vibrated granular system confined to a quasi-two-dimensional geometry. In this setup, the particles gain kinetic energy through their collisions with the two horizontal walls, separated a distance smaller than twice the particle diameter. This energy is transferred to the horizontal degrees of freedom by means of  the (inelastic) collisions between particles. A steady state is reached when the two mechanisms, energy dissipation because of the inelasticity of collisions and energy injection due to the energy transferred from the vertical to the horizontal motion, cancel each other. The idea of the model \cite{BRyS13} is to introduce some modification of the collision rule, trying to describe the mechanism for which the kinetic energy associated to the motion in the vertical direction is transformed into kinetic energy in the horizontal plane. In other words, an effective horizontal dynamics, in which the effects of the vertical injection of energy is taken into account in the modified, effective collision rule, is proposed. The model was designed trying to mimic what had been observed in experiments of quasi-two-dimensional vibrated granular systems \cite{RPGRSCyM11,OyU05}. What renders the model quite appealing and  interesting is that, contrary to models based on stochastic thermostats, the physical mechanism leading to the effective dynamics is now well identified. The vertical dynamics occurs over a much faster time scale than the horizontal one. Moreover the latter is characterized by momentum and density conservation. This detailed formulation of the model opens the possibility of deriving such effective confined dynamics starting from the the Newton description in three dimensions. Since the model is formulated by means of a collision rule, it is possible to develop a self-consistent description at the level of both statistical mechanics and kinetic theory, by  directly extending the methods already developed for inelastic hard spheres and disks \cite{BDyS97,vNEyB98}.

In this paper, the binary collision operators describing the particle dynamics introduced in ref. \cite{BRyS13} are identified and, afterwards, the pseudo-Liouville equation governing the time evolution of the $N$-particle distribution function of the system is constructed. Then the existence of a steady state and the scaling properties of its distribution function are analyzed. To allow explicit calculations, the Enskog approximation is discussed and used as the basis for the derivation of the temperature of the steady state and also of the one-particle velocity distribution, in some mathematically well defined approximation. The theoretical predictions are compared with the results obtained by means of numerical simulations, and a good agreement is observed. This puts the theory developed on a firm basis and opens the possibility of deriving macroscopic transport equations  with explicit expressions for the involved coefficients, valid beyond the quasi-elastic limit.

The organization of the remaining of the paper is as follows. The collision rule and some of its implications are discussed in Sec. \ref{s2}, while in Sec.\ \ref{s3} the formal statistical description of the system is developed. Also the Enskog approximation is introduced and discussed in some detail. Special emphasis is put on formulating the hypothesis leading to the Enskog theory in a concise way and on the fact that it accounts for the existence of non-equilibrium velocity correlations. Section \ref{s4} is devoted to the study of the steady state eventually reached by the system. Some exact scaling of its distribution function is identified and its consequences on the properties of the steady temperature are pointed out. Approximated explicit expressions for the latter and also for the one-particle distribution function of the steady state are derived in the framework of the Enskog theory. In Sec. \ref{s5}, the predictions are compared with numerical results obtained by numerical simulation, namely the direct simulation Monte Carlo method \cite{Bi94,Ga00}. The paper concludes with a short summary and some comments on the results presented.

\section{Description of the model}
\label{s2}
Consider a system composed of $N$ smooth inelastic hard spheres ($d=3$) or disks ($d=2$) of mass $m$ and diameter $\sigma$ confined in a volume $V$. The position and velocity of the $i$-th particle will be denoted by ${\bm r}_{i}$ and ${\bm v}_{i}$, respectively, while ${\bm x}_{i}$ will be used for the combined variable including both position and velocity, i.e. ${\bm x}_{i} \equiv \left\{ {\bm r}_{i}, {\bm v}_{i} \right\}$. The dynamics of the system consists of free streaming of the particles interrupted by instantaneous binary collisions. In a collision, the velocities ${\bm v}_{1}$ and ${\bm v}_{2}$ of the two involved particles change according to the deterministic rule \cite{BRyS13}
\begin{equation}
\label{2.1}
{\bm v}_{1} \rightarrow {\bm v}^{\prime}_{1} = {\bm v}_{1}- \frac{1+ \alpha}{2}  {\bm v}_{12} \cdot \widehat{\bm \sigma}  \widehat{\bm \sigma} + \Delta \widehat{\bm \sigma},
\end{equation}
\begin{equation}
\label{2.2}
{\bm v}_{2} \rightarrow {\bm v}^{\prime}_{2} = {\bm v}_{2}+ \frac{1+ \alpha}{2}  {\bm v}_{12} \cdot \widehat{\bm \sigma} \widehat{\bm \sigma} - \Delta \widehat{\bm \sigma},
\end{equation}
where ${\bm v}_{12} \equiv {\bm v}_{1}-{\bm v}_{2}$ is the relative velocity prior to collision, $\widehat{\bm \sigma}$ is the unit vector pointing from the center of particle $2$ to the center of particle $1$ at contact, and $\Delta$ is a positive characteristic constant velocity. Finally, $\alpha$ is the coefficient of normal restitution, defined in the interval $ 0 < \alpha \leq 1$. The above scattering law conserves total lineal momentum and also angular momentum. The term proportional to $\Delta$ in the collision rule tries to describe the mechanism for which the energy given to some (irrelevant) degrees of freedom of the particles is transferred to the other (relevant) degrees of freedom. Although in the original formulation of the model the irrelevant motion was the vertical one and, therefore, the relevant degrees of freedom correspond to the motion of the particles in the horizontal plane \cite{BRyS13}, the most general case in which the effective motion can occur in the three-dimensional space will be considered here.

The relative velocity after the collision is
\begin{equation}
\label{2.3}
{\bm v}^{\prime}_{12}\equiv {\bm v}^{\prime}_{1}-{\bm v}^{\prime}_{2} = {\bm v}_{12}-(1+\alpha)  {\bm v}_{12} \cdot \widehat{\bm \sigma} \widehat{\bm \sigma} + 2 \Delta \widehat{\bm \sigma},
\end{equation}
so that
\begin{equation}
\label{2.4}
{\bm v}_{12}^{\prime} \cdot \widehat{\bm \sigma} =-\alpha {\bm v}_{12} \cdot \widehat{\bm \sigma} + 2 \Delta.
\end{equation}
For a collision to happen, it must be  ${\bm v}_{12} \cdot \widehat{\bm \sigma}  <0$ and, therefore
\begin{equation}
\label{2.5}
| {\bm v}_{12}^{\prime} \cdot \widehat{\bm \sigma} |= \alpha |{\bm v}_{12} \cdot \widehat{\bm \sigma}| + 2 \Delta.
\end{equation}
The change in kinetic energy upon collision is
\begin{equation}
\label{2.6}
e^{\prime}-e \equiv \frac{m}{2} \left(v^{\prime 2}_{1}+ v^{\prime 2}_{2}-v_{1}^{2}-v_{2}^{2} \right) = m\left[ \Delta^{2} - \alpha \Delta  {\bm v}_{12} \cdot \widehat{\bm \sigma} - \frac{1-\alpha^{2}}{4}\,  ({\bm v}_{12} \cdot \widehat{\bm \sigma} )^{2} \right].
\end{equation}
It follows that energy can be gained or lost in a collision depending on whether $|{\bm v}_{12} \cdot \widehat{\bm \sigma} |$ is smaller or larger than $2 \Delta /(1-\alpha)$.

It is convenient to consider also the restitution collision corresponding to Eqs.\ (\ref{2.1}) and (\ref{2.2}). It is defined as the collision leading to the after collision velocities ${\bm v}_{1}$ and ${\bm v}_{2}$ with the same collision vector, $\widehat{\bm \sigma}$. The pre-collisional velocities ${\bm v}^{*}_{1}$ and ${\bm v}^{*}_{2}$ for the restitution collision are
\begin{equation}
\label{2.7} {\bm v}^{*}_{1} = {\bm v}_{1} - \frac{1+ \alpha}{2 \alpha}\,  {\bm v}_{12} \cdot \widehat{\bm \sigma}  \widehat{\bm \sigma} + \frac{\Delta \widehat{\bm \sigma}}{\alpha},
\end{equation}
\begin{equation}
\label{2.8} {\bm v}^{*}_{2} = {\bm v}_{2} + \frac{1+ \alpha}{2 \alpha}\, {\bm v}_{12} \cdot \widehat{\bm \sigma}  \widehat{\bm \sigma} - \frac{\Delta \widehat{\bm \sigma}}{\alpha}\, .
\end{equation}
For the variation of kinetic energy, it is
\begin{equation}
\label{2.9}
e^{*}-e= m \left[ \frac{\Delta^{2}}{\alpha^{2}}+ \frac{1-\alpha^{2}}{4 \alpha^{2}}\, \left(  {\bm v}_{12} \cdot \widehat{\bm \sigma}  \right)^{2} - \frac{\Delta}{\alpha^{2}}\, {\bm v}_{12} \cdot \widehat{\bm \sigma} \right].
\end{equation}
The volume transformation in velocity space in a collision is
\begin{equation}
\label{2.10}
d{\bm v}^{\prime}_{1} d {\bm v}^{\prime}_{2} = \alpha d{\bm v}_{1} d{\bm v}_{2},
\end{equation}
and, consequently,
\begin{equation}
\label{2.11}
d{\bm v}^{*}_{1} d {\bm v}^{*}_{2} = \alpha^{-1}  d{\bm v}_{1} d{\bm v}_{2}.
\end{equation}
This is the same result as for smooth inelastic hard spheres or disks. The mathematical reason for this is that the terms proportional to $\Delta$ in the collision rule do not depend on the pre-collisional velocities of the particles.

This completes the mechanical specification of the model, which in the following will be referred to as a system of repulsive inelastic hard spheres or disks.

\section{Pseudo-Liouville operators and the Enskog theory}
\label{s3}
The conditions leading to a collision of two hard spheres  or disks, i.e. the identification of the collision cylinder, are independent from the collision rule. As a consequence, the generator, $L_{+}(\Gamma)$, for the dynamics of a phase space function $A(\Gamma)$, $\Gamma \equiv \left\{ {\bm x}_{1}, {\bm x}_{2}, \ldots, {\bm x}_{N} \right\}$, follows directly by analogy with the elastic fluid of hard spheres or disks \cite{EDHyvL69,BDyS97}. It is defined by
\begin{equation}
\label{3.1}
A({\Gamma},t) \equiv A \left[ \Gamma (t) \right] \equiv e^{t L_{+}(\Gamma)} A(\Gamma),
\end{equation}
where $\Gamma (t)$ is the phase point at time $t$ according to the dynamics of the system, given that it was initially at $\Gamma$. Actually, it is enough to define
\begin{equation}
\label{3.2}
W(\left\{ {\bm r}_{i} \right\})  e^{t L_{+}(\Gamma)} A(\Gamma),
\end{equation}
with $W(\left\{ {\bm r}_{i} \right\})$ being an overlap function which vanishes for any configuration of the system having two overlapping particles and it is unity otherwise,
\begin{equation}
\label{3.2a}
W(\left\{ {\bm r}_{i} \right\})  = \prod_{1 \leq i < j \leq N} \theta (r_{ij}- \sigma).
\end{equation}
Here $\theta (x)$ is the Heaviside step function defined as $\theta (x) = 1$ for $x\geq 0$ and $\theta (x) =0$ for $x < 0$.
The generator reads \cite{EDHyvL69,BDyS97}
\begin{equation}
\label{3.3}
L_{+}(\Gamma) = \sum_{i=1}^{N} {\bm v}_{i} \cdot \frac{\partial}{\partial {\bm r}_{i}}+ \sum_{1 \leq i < j \leq N} T_{+}({\bm x}_{i},{\bm x}_{j}).
\end{equation}
The first term on the right hand side generates free streaming while the second one describes instantaneous velocity changes in collisions. The binary collision operator $T_{+}({\bm x}_{i},{\bm x}_{j})$ for particles $i$ and $j$ is given by
\begin{equation}
\label{3.4}
T_{+}({\bm x}_{i},{\bm x}_{j}) = \sigma^{d-1} \int d \widehat{\bm \sigma}\, \theta (- {\bm v}_{12} \cdot \widehat{\bm \sigma} ) | {\bm v}_{12} \cdot \widehat{\bm \sigma}  |  \delta ({\bm r}_{ij}-{\bm \sigma}) \left[ b_{\bm \sigma}(i,j)-1 \right].
\end{equation}
 In the above expression, $d \widehat{\bm \sigma}$ denotes the solid angle element for $\widehat{\bm \sigma}$, ${\bm \sigma}= \sigma \widehat{\bm \sigma}$, ${\bm r}_{ij} \equiv {\bm r}_{i}-{\bm r}_{j}$ is the relative position vector of the two particles, and the operator $b_{\bm \sigma}(i,j)$ replaces all the velocities ${\bm v}_{i}$ and ${\bm v}_{j}$ to its right by their post-collisional values, obtained with the collision rule given in Eqs.\ (\ref{2.1}) and (\ref{2.2}). This is the only point in which the details of the collision rule show up. Otherwise, the form of $L_{+}(\Gamma)$ is the same for all kind of smooth hard spheres or disks.

To express the dynamics of the system in terms of the evolution of the probability density, instead of the evolution of the phase space functions, another generator $\overline{L}_{+}(\Gamma)$ is defined as
\begin{equation}
\label{3.5}
\int d\Gamma\, B(\Gamma) W(\{ {\bm r}_{i}\}) L_{+}(\Gamma) A(\Gamma)= \int d\Gamma \left[ \overline{L}_{+}(\Gamma) W(\{{\bm r}_{i} \}) B(\Gamma) \right] A(\Gamma),
\end{equation}
where $d\Gamma \equiv d{\bm x}_{1}...d{\bm x}_{N}$, for arbitrary functions $A(\Gamma)$ and $B(\Gamma)$. It is easily found that
\begin{equation}
\label{3.6}
\overline{L}_{+} (\Gamma)  \equiv - \sum_{i=1}^{N} {\bm v}_{i} \cdot \frac{\partial}{\partial {\bm r}_{i}} +
\sum_{1 \leq i < j \leq N}\overline{T}_{+}({\bm x}_{i},{\bm x}_{j}),
\end{equation}
with the new binary collision operator,
\begin{eqnarray}
\label{3.7}
\overline{T}_{+}({\bm x}_{i},{\bm x}_{j}) & = &  \sigma^{d-1} \int d \widehat{\bm \sigma}\, \left[ \theta({\bm v}_{ij} \cdot \widehat{\bm \sigma} - 2 \Delta) ({\bm v}_{ij} \cdot \widehat{\bm \sigma}  - 2 \Delta) \delta ({\bm r}_{ij} -{\bm \sigma}) \alpha^{-2} b_{\bm \sigma}^{-1} (i,j) \right. \nonumber \\
& & - \left. \theta ( {\bm v}_{ij} \cdot \widehat{\bm \sigma} ){\bm v}_{ij} \cdot \widehat{\bm \sigma}  \delta ({\bm r}_{ij} + {\bm \sigma}) \right].
\end{eqnarray}
The operator $b_{\bm \sigma}^{-1}(i,j)$ is the inverse of $b_{\bm \sigma}(i,j)$, i.e. it changes all the velocities ${\bm v}_{i}$ and ${\bm v}_{j}$ to its right into the pre-collisional values ${\bm v}^{*}_{1}$ and ${\bm v}^{*}_{2}$ given by Eqs.\ (\ref{2.7}) and (\ref{2.8}). The two binary collision operators $T_{+}$ and $\overline{T}_{+}$ verify
the relationship
\begin{equation}
\label{3.8}
\int d{\bm v}_{i} \int d{\bm v}_{j}\, B({\bm x}_{i},{\bm x}_{j}) T_{+}({\bm x}_{i},{\bm x}_{j}) A({\bm x}_{i},{\bm x}_{j}) =\int d{\bm v}_{i} \int d{\bm v}_{j}\, A({\bm x}_{i},{\bm x}_{j}) \overline{T}_{+}({\bm x}_{i},{\bm x}_{j}) B({\bm x}_{i},{\bm x}_{j}),
\end{equation}
for arbitrary $A$ an $B$. The average value of a phase function $A(\Gamma)$ at time $t$ is
\begin{equation}
\label{3.9}
<A(t)> = \int d \Gamma\, \rho (\Gamma) A(\Gamma,t)= \int d \Gamma\, \rho(\Gamma) e^{t L_{+}(\Gamma)} A(\Gamma),
\end{equation}
where $\rho(\Gamma)$ is the probability distribution of initial conditions, giving a vanishing probability to overlapping configurations, i.e. it has an intrinsic $W(\{ {\bm r}_{i} \})$ factor. Using the definition of $\overline{L}_{+}(\Gamma)$ given in Eq.\ (\ref{3.5}), the above average can be written as
\begin{equation}
\label{3.10}
<A(t)>= \int d\Gamma\ \rho(\Gamma,t) A(\Gamma),
\end{equation}
with
\begin{equation}
\label{3.11}
\rho(\Gamma,t) \equiv e^{t \overline{L}_{+}(\Gamma)} \rho (\Gamma).
\end{equation}
This expression defines the time dependence of the probability distribution density $\rho(\Gamma,t)$, which therefore evolves in time according to the pseudo-Liouville equation
\begin{equation}
\label{3.12}
\left[ \frac{\partial}{\partial t} - \overline{L}_{+}(\Gamma) \right] \rho(\Gamma,t)= 0.
\end{equation}
Reduced distribution functions of $l$ particles, $f_{l}({\bm x}_{1}, \cdots, {\bm x}_{l},t)$, are defined as
\begin{equation}
\label{3.13}
f_{l}({\bm x}_{1}, \cdots, {\bm x}_{l},t)= \frac{N!}{(N-l)!} \int d{\bm x}_{l+1} \ldots d{\bm x}_{N}\, \rho (\Gamma,t).
\end{equation}
These functions obey the Born-Bogoliubov-Green-Kirkwood-Yvon (BBGKY) hierarchy \cite{McL89}, which follows by partial integration of the pseudo-Liouville equation over the phase space variables ${\bm x}_{l+1}, \cdots, {\bm x}_{N}$,
\begin{equation}
\label{3.14}
\left[ \frac{\partial}{\partial t} - \overline{L}_{+}({\bm x}_{1}, \cdots, {\bm x}_{l}) \right] f_{l}({\bm x}_{1},\cdots, {\bm x}_{l},t) = \sum_{i=1}^{l} \int d{\bm x}_{l+1}\, \overline{T}_{+}({\bm x}_{i},{\bm x}_{l+1}) f_{l+1} ({\bm x}_{1}, \cdots, {\bm x}_{l+1},t),
\end{equation}
where $\overline{L}_{+}({\bm x}_{1}, \cdots, {\bm x}_{l})$ is the generator of the dynamics for a system of $l$ particles
\begin{equation}
\label{3.15}
\overline{L}_{+} ({\bm x}_{1}, \cdots,{\bm x}_{l})  \equiv - \sum_{i=1}^{l} {\bm v}_{i} \cdot \frac{\partial}{\partial {\bm r}_{i}} + \sum_{ 1 \leq i < j \leq N }\overline{T}_{+}({\bm x}_{i},{\bm x}_{j}).
\end{equation}
In particular, the first equation of the BBGKY hierarchy reads
\begin{equation}
\label{3.16}
\left( \frac{\partial}{\partial t} +{\bm v}_{1} \cdot \frac{\partial}{\partial {\bm r}_{1}} \right) f_{1}({\bm x}_{1},t) = \int d{\bm x}_{2}\, \overline{T}_{+}({\bm x}_{1}, {\bm x}_{2} ) f_{2}({\bm x}_{1},{\bm x}_{2},t).
\end{equation}
A formal kinetic equation for the one-particle distribution function, $f_{1}({\bm x}_{1},t)$, is obtained if $f_{2}({\bm x}_{1},{\bm x}_{2},t)$ is expressed in the right hand side of the above equation as some functional of $f_{1}({\bm x}_{1},t)$. The most common approximation is to neglect the correlations of the velocities of the two colliding particles, prior to the collision. To see that only the pre-collisional part of the two-body reduced distribution is needed in Eq.\ (\ref{3.16}), note the relation
\begin{equation}
\label{3.17}
\theta (\widehat{\bm \sigma} \cdot {\bm v}_{ij} - 2 \Delta ) b_{\bm \sigma}^{-1} (i,j)= b_{\bm \sigma}^{-1} (i,j) \theta (-\widehat{\bm \sigma} \cdot {\bm v}_{ij}),
\end{equation}
so that Eq.\ (\ref{3.7}) can be rewritten in the equivalent form
\begin{eqnarray}
\label{3.18}
\overline{T}_{+}({\bm x}_{i},{\bm x}_{j}) & = &  \sigma^{d-1} \int d \widehat{\bm \sigma}\, \delta ({\bm r}_{ij} -{\bm \sigma}) \left[ |{\bm v}_{ij}  \cdot \widehat{\bm \sigma}- 2 \Delta | \alpha^{-2} b_{\bm \sigma}^{-1} (i,j)-|  {\bm v}_{ij} \cdot \widehat{\bm \sigma}| \right] \theta (- {\bm v}_{ij} \cdot \widehat{\bm \sigma})  \nonumber \\
&=& \delta (r_{ij}-\sigma) \left[ | {\bm v}_{ij} \cdot \widehat{\bm r}_{ij} - 2\Delta | \alpha^{-2} b_{\bm r}^{-1} (i,j) - | {\bm v}_{ij} \cdot \widehat{\bm r}_{ij} | \right] \theta(- {\bm v}_{ij} \cdot \widehat{\bm r}_{ij}),
\end{eqnarray}
where $\widehat{\bm r}_{ij} \equiv {\bm r}_{ij}/r_{ij}$. Upon writing the first equality, $\widehat{\bm \sigma}$ has been changed into $-\widehat{\bm \sigma}$ in the second term under the integral before carrying out the angular integration. The step function $\theta ( - {\bm v}_{ij} \cdot \widehat{\bm r}_{ij})$ and the factor $\delta(r_{ij}-\sigma)$ restrict the needed information to particles at contact before collision.

In the Enskog approximation, the kinetic equation is derived by assuming that
\begin{equation}
\label{3.19}
\delta(r_{12}-\sigma) \theta (- {\bm v}_{12}  \cdot \widehat{\bm r}_{12}) f_{2} ({\bm x}_{1},{\bm x}_{2},t)\approx \delta (r_{12} -\sigma) \theta (- {\bm v}_{12} \cdot \widehat{\bm r}_{12} ) g({\bm r}_{1},{\bm r}_{2},t) f_{1}({\bm x}_{1},t) f_{1}({\bm x}_{2},t).
\end{equation}
The factor $g({\bm r}_{1},{\bm r}_{2},t)$ is the spacial pair correlation function. In the revised Enskog theory (RET) \cite{vByE79}, this quantity is approximated by the equilibrium functional of the density, $g_{E}$,  evaluated with the non-equilibrium density field at time $t$. Substitution of Eq.\ (\ref{3.19}) into Eq.\ (\ref{3.16}) gives the RET, generalized to the present collision rule,
\begin{equation}
\label{3.20}
\left( \frac{\partial}{\partial t} +{\bm v}_{1} \cdot \frac{\partial}{\partial {\bm r}_{1}} \right) f_{1}({\bm x}_{1},t) = \int d{\bm x}_{2}\, \overline{T}_{+}({\bm x}_{1}, {\bm x}_{2} ) g_{E}[{\bm r}_{1},{\bm r}_{2} |n(t)] f({\bm x}_{1},t)f_{1}({\bm x}_{2},t).
\end{equation}
It is worth to stress that the RET provides a description of the dynamics over the whole range of densities and length scales, including both fluid and cristal phases \cite{KDEyP90}.

The Enskog approximation also has consequences on the correlations \cite{Lu99}. The second equation of the BBGKY hierarchy, Eq.\  (\ref{3.14}), reads
\begin{eqnarray}
\label{3.21}
\left ( \frac{\partial}{\partial t} +{\bm v}_{1} \cdot \frac{\partial}{\partial {\bm r}_{1}} + {\bm v}_{2} \cdot \frac{\partial}{\partial {\bm r}_{2 }} \right)&& f_{2} ({\bm x}_{1},{\bm x}_{2},t) = \overline{T}_{+}({\bm x}_{1},{\bm x}_{2})  f_{2}({\bm x}_{1},{\bm x}_{2},t)  \nonumber \\
&& +  \int d{\bm x}_{3}\, \left[ \overline{T}_{+} ({\bm x}_{1},{\bm x}_{3} ) + \overline{T}_{+}({\bm x}_{2},{\bm x}_{3}) \right] f_{3}({\bm x}_{1},{\bm x}_{2},{\bm x}_{3},t).
\end{eqnarray}
For physically relevant initial conditions in which particles do not overlap, it is
\begin{equation}
\label{3.22}
f_{s}({\bm x}_{1}, \ldots,{\bm x}_{s})= W ({\bm r}_{1},\ldots, {\bm r}_{s}) f_{s0}({\bm x}_{1}, \ldots, {\bm x}_{s}),
\end{equation}
where $W({\bm r}_{1}, \ldots,{\bm r}_{s})$ is the overlap function, defined in Eq\ (\ref{3.2a}) for the $s$ particles. Of course, Eq.\ (\ref{3.22}) does not define the function $f_{s0}$ in  unique way. Without restriction, it will be assumed that it is regular everywhere as well as its derivatives. Then, Eq.\ (\ref{3.21}) can be decomposed into two separate equations, containing only regular and singular terms at $r_{12}= \sigma$, respectively. The equation with the singular contributions reads
\begin{equation}
\label{3.23}
f_{2}({\bm x}_{1},{\bm x}_{2},t) {\bm v}_{12} \cdot \widehat{\bm r}_{12} \delta (r_{12} - \sigma) - \overline{T}_{+} ({\bm x}_{1},{\bm x}_{2} ) f_{2} ({\bm x}_{1},{\bm x}_{2},t)=0
\end{equation}
and by means of the RET it is found after some rearrangements that the pair correlation function at contact can be expressed as
\begin{eqnarray}
\label{3.24}
g({\bm r}_{1},{\bm r}_{2},t) \delta ( r_{12}- \sigma)& = & g_{E}({\bm r}_{1},{\bm r}_{2},t) \delta ( r_{12} -\sigma)
+ \frac{1}{n_{1}({\bm r}_{1},t) n({\bm r}_{2},t)} \int d{\bm v}_{1} \int d{\bm v}_{2} \nonumber \\
&& \times \left[  \frac{| {\bm v}_{12} \cdot \widehat{\bm r}_{12} |}{ 2 \Delta + \alpha | {\bm v}_{12} \cdot \widehat{\bm r}_{12}|}\,  \theta (-{\bm v}_{12} \cdot \widehat{\bm r}_{12} ) - \theta ({\bm v}_{12} \cdot \widehat{\bm r}_{12} )\right] \nonumber \\
&& \times f({\bm x}_{1},t) f({\bm x}_{2},t) g_{E}({\bm r}_{1},{\bm r}_{2},t) \delta (r_{12}-\sigma).
\end{eqnarray}
Some details of the calculations are given in the Appendix. This result expresses the two-body correlation  function at contact as the sum of two rather different terms. The first contribution only contains space correlations as described by the equilibrium pair correlation functional. The second term on the right hand side of Eq.\ (\ref{3.24}) is a correction taking into account the velocity correlations generated by the collision that, in turn, generate position correlations.

In the low density limit, the RET leads to the Boltzmann kinetic theory. Then putting $g_{E}=1$ in Eq.\ (\ref{3.20}) the Boltzmann equation for the model follows. The same substitution in Eq.\ (\ref{3.24}) provides the expression of the pair correlation at contact in the low density limit. It is worth to stress that this correlations do not vanish in general for non-equilibrium systems.

\section{The homogeneous steady state}
\label{s4}
Introduce dimensionless space and time scales defined by
\begin{equation}
\label{4.1}
{\bm q}_{i} \equiv \frac{{\bm r}_{i}}{\ell},
\end{equation}
\and
\begin{equation}
\label{4.2}
\tau \equiv \frac{\Delta}{\ell}\, t,
\end{equation}
respectively. Here $\ell \equiv \left( n \sigma^{d-1} \right)^{-1}$, with $n \equiv N/V$ being the average density. Consistently with the above, the dimensionless velocities are
\begin{equation}
\label{4.3}
{\bm \omega}_{i} \equiv \frac{{\bm v}_{i}}{\Delta}.
\end{equation}
The distribution function $\widetilde{\rho}$ in the new phase space, $\widetilde{\Gamma} \equiv \{ \widetilde{\bm x}_{1}, \ldots, \widetilde{\bm x}_{N} \}$, $\widetilde{\bm x}_{i} \equiv \{ {\bm q}_{i},{\bm \omega}_{i} \}$, is related with the old one by
\begin{equation}
\label{4.4}
\widetilde{\rho} (\widetilde{\Gamma},\tau) = \left( \ell \Delta \right)^{d N} \rho (\Gamma,t).
\end{equation}
Substitution of this into the pseudo-Liouville equation (\ref{3.12}) yields
\begin{equation}
\label{4.5}
\left[ \frac{\partial}{\partial \tau} - \widetilde{\overline{L}}_{+}(\widetilde{\Gamma}) \right] \widetilde{\rho} (\widetilde{\Gamma}, \tau)=0,
\end{equation}
with the definitions
\begin{equation}
\label{4.6}
\widetilde{\overline{L}} (\widetilde{\Gamma}) = -\sum_{i=1}^{N}{\bm \omega}_{i} \cdot \frac{\partial}{\partial {\bm q}_{i}}+  \sum_{1 \leq i <j \leq N} \widetilde{\overline{T}} (\widetilde{\bm x}_{i}, \widetilde{\bm x}_{j}),
\end{equation}
\begin{eqnarray}
\label{4.7}
\widetilde{\overline{T}}_{+}(\widetilde{\bm x}_{i},\widetilde{\bm x}_{j}) & = &  \widetilde{\sigma}^{d-1} \int d \widehat{\bm \sigma}\, \left[ \theta({\bm \omega}_{ij} \cdot \widehat{\bm \sigma} - 2 ) ({\bm \omega}_{ij} \cdot \widehat{\bm \sigma}  - 2 ) \delta ({\bm q}_{ij} -\widetilde{\bm \sigma}) \alpha^{-2} b_{\bm \sigma}^{-1} (i,j) \right. \nonumber \\
& & - \left. \theta ( {\bm \omega}_{ij} \cdot \widehat{\bm \sigma} ){\bm \omega}_{ij} \cdot \widehat{\bm \sigma}  \delta ({\bm q}_{ij} + \widetilde{\bm \sigma}) \right].
\end{eqnarray}
where $\widetilde{\sigma} \equiv \sigma / \ell$, $\widetilde{\bm \sigma} \equiv {\bm \sigma}/\ell$, and the operator $b_{\bm \sigma}^{-1} (i,j)$ now acts on the velocities ${\bm \omega}_{i}$ and ${\bm \omega}_{j}$, changing them into
\begin{equation}
\label{4.8}
b_{\sigma}^{-1}(i,j) {\bm \omega}_{i} = {\bm \omega}_{i}^{*} = {\bm \omega}_{i}- \frac{1+\alpha}{2 \alpha} {\bm \omega}_{ij} \cdot \widehat{\bm \sigma} \widehat{\bm \sigma} +\frac{\bm \sigma}{\alpha}\, ,
\end{equation}
\begin{equation}
\label{4.9}
b_{\sigma}^{-1}(i,j) {\bm \omega}_{j} = {\bm \omega}_{j}^{*} = {\bm \omega}_{j}+ \frac{1+\alpha}{2 \alpha} {\bm \omega}_{ij} \cdot \widehat{\bm \sigma} \widehat{\bm \sigma} -\frac{\bm \sigma}{\alpha}\, .
\end{equation}
The relevant issue is that all the dependence on $\Delta$ has been scaled out. The dynamics of the system in the phase space $\widetilde{\Gamma}$ and in the time scale $\tau$ does not depend on the value of $\Delta$. This property has relevant implications. Consider a steady state. Its distribution function has the form,
\begin{equation}
\label{4.10}
\rho_{s} (\Gamma)= \left( \ell \Delta \right)^{-dN} \widetilde{\rho}_{s} (\widetilde{\Gamma}),
\end{equation}
where $\widetilde{\rho}_{s}(\widetilde{\Gamma})$ is a steady solution of Eq.\  (\ref{4.5}) and, therefore, independent ´from $\Delta$ (although depending on the coefficient of normal restitution $\alpha$). The steady average of a dynamic variable $A(\Gamma)$ is
\begin{equation}
\label{4.11}
\langle A \rangle_{s}= \int d \Gamma\, A(\Gamma) \rho_{s}(\Gamma) = \int d \widetilde{\Gamma} A(\left\{ \ell {\bm q}_{i} \right\}, \left\{ \Delta {\bm \omega}_{i} \right\} ) \widetilde{\rho}_{s} (\widetilde{\Gamma}).
\end{equation}
If $A$ is an homogeneous function of degree $a$ of the velocity,
\begin{equation}
\label{4.12}
A(\left\{ \ell {\bm q}_{i} \right\}, \left\{ \Delta {\bm \omega}_{i} \right\} ) = \Delta^{a} A(\left\{ \ell {\bm q}_{i} \right\}, \left\{  {\bm \omega}_{i} \right\} ),
\end{equation}
Eq.\ (\ref{4.11}) yields
\begin{equation}
\label{4.13}
\langle A \rangle_{s}= \Delta^{a} \int d \widetilde{\Gamma} A(\left\{ \ell {\bm q}_{i} \right\}, \left\{ {\bm \omega}_{i} \right\} ) \widetilde{\rho}_{s} (\widetilde{\Gamma}).
\end{equation}
The dependence on $\Delta$ of the average is trivially identified. As a  prototypical application of the above, consider the granular temperature of the system $T(t)$, which for translationally invariant states is defined in terms of the average kinetic energy density $E_{K}(\Gamma)$ as
\begin{equation}
\label{4.14}
\frac{d N T(t)}{2} = < E_{K}(t)>,
\end{equation}
\begin{equation}
\label{4.15}
E_{K}(\Gamma) \equiv \sum_{i=1}^{N} \frac{m v_{i}^{2}}{2}.
\end{equation}
As usual in the granular matter literature, the Boltzmann constant has been set formally equal to unity. Then, the temperature $T_{s}$ of the steady state can be written as
\begin{equation}
\label{4.16}
\frac{d N T_{s}}{2} = \Delta^{2} \sum_{i=1}^{N} \int d \widetilde{\Gamma}\, \frac{m \omega_{i}^{2}}{2} \widetilde{\rho}_{s} (\widetilde{\Gamma}).
\end{equation}
Consequently, $T_{s}/\Delta^{2}$ is a function of $\alpha$, being independent from $\Delta$. The scaling property in Eq.\ (\ref{4.10}) translates to the reduced distribution functions of the steady state,
\begin{equation}
\label{4.17}
f_{l,s}({\bm x}_{1}, \cdots, {\bm x}_{l})= \left( \ell \Delta \right)^{-l d} \widetilde{f}_{l,s} (\widetilde{\bm x}_{1}, \cdots \widetilde{\bm x}_{l}),
\end{equation}
with
\begin{equation}
\label{4.18}
\widetilde{f}_{l,s} (\widetilde{\bm x}_{1}, \cdots \widetilde{\bm x}_{l}) \equiv \frac{N!}{(N-l)!} \int d \widetilde{\bm x}_{l+1} \ldots d \widetilde{\bm x}_{N} \widetilde{\rho}_{s} (\widetilde{\Gamma}).
\end{equation}
These reduced distribution functions do not depend on the velocity parameter $\Delta$. The expression of the steady temperature in Eq. (\ref{4.16}) can be expressed in the equivalent form,
\begin{equation}
\label{4.19}
T_{s}= \frac{m \Delta^{2}}{\widetilde{n}d} \int d{\bm \omega}_{1} \omega_{1}^{2} \widetilde{f}_{1,s} (\omega_{1}),
\end{equation}
where $\widetilde{n} \equiv n \ell^{d}$ and it has been used that the homogeneity and isotropy of the steady state implies that the one-particle distribution function can not depend on the position or the direction of the velocity. An expression for the steady temperature can be derived as follows. Stationarity of $\widetilde{\rho}_{s}$ yields
\begin{equation}
\label{4.20}
\sum_{i=1}^{N} \int d \widetilde{\Gamma}\, \frac{m \omega_{i}^{2}}{2}\, \widetilde{\overline{L}}_{+} (\widetilde{\Gamma}) \widetilde{\rho}_{s}(\widetilde{\Gamma})=0.
\end{equation}
Neglecting surface contributions, this is seen to be equivalent to
\begin{equation}
\label{4.21}
\int d \widetilde{\bm x}_{1} \int d \widetilde{\bm x}_{2}\, \left\{ \widetilde{T}_{+}(\widetilde{\bm x}_{1}, \widetilde{\bm x}_{2}) \left[ \frac{m}{2}\, (\omega_{1}^{2} + \omega_{2}^{2} ) \right] \right\} \widetilde{f}_{2,s} (\widetilde{\bm x}_{1}, \widetilde{\bm x}_{2})=0.
\end{equation}
In the above expression, $\widetilde{T}_{+}$ is the binary collision operator,
\begin{equation}
\label{4.22}
\widetilde{T}_{+}(\widetilde{\bm x}_{1},\widetilde{\bm x}_{2}) = \widetilde{\sigma}^{d-1} \int d \widehat{\bm \sigma}\, \theta (- {\bm \omega}_{12} \cdot \widehat{\bm \sigma} ) | {\bm \omega}_{12} \cdot \widehat{\bm \sigma}  |  \delta ({\bm q}_{12}-\widetilde{\bm \sigma}) \left[ b_{\bm \sigma}(1,2)-1 \right],
\end{equation}
\begin{equation}
\label{4.23}
b_{\bm \sigma}(1,2) {\bm \omega}_{1} = {\bm \omega}_{1}^{\prime} = {\bm \omega}_{1}- \frac{1+\alpha}{2 } {\bm \omega}_{12} \cdot \widehat{\bm \sigma} \widehat{\bm \sigma} +{\bm \sigma} ,
\end{equation}
\begin{equation}
\label{4.24}
b_{\bm \sigma}(1,2) {\bm \omega}_{2} = {\bm \omega}_{2}^{\prime} = {\bm \omega}_{2}+ \frac{1+\alpha}{2 } {\bm \omega}_{12} \cdot \widehat{\bm \sigma} \widehat{\bm \sigma} -{\bm \sigma} .
\end{equation}
In the Enskog theory discussed in Sec. \ref{s3}, the pre-collisional two body distribution function in Eq.\ (\ref{4.21}) is approximated by
\begin{equation}
\label{4.25}
\delta ({\bm q}_{12} - \widetilde{\bm \sigma} ) \widetilde{f}_{2,s} (\widetilde{\bm x}_{1}, \widetilde{\bm x}_{2}) \approx \widetilde{n}^{2} \phi(\omega_{1}) \phi (\omega_{2}) g_{e}(\sigma;n) \delta ({\bm q}_{12} - \widetilde{\bm \sigma} ),
\end{equation}
where $g_{e}(\sigma;n)$ is the equilibrium pair correlation function at contact and $\phi(\omega)$ is defined by
\begin{equation}
\label{4.26}
\widetilde{f}_{1,s} (\widetilde{\bm x}) = \widetilde{n}  \phi (\omega).
\end{equation}
When Eq.\ (\ref{4.25}) is used into Eq.\ (\ref{4.21}) and the angular integrations are carried out, it is obtained:
\begin{equation}
\label{4.27}
\int d \widetilde{\bm \omega}_{1} \int d \widetilde{\bm \omega}_{2}\, \left[ \frac{ \omega_{12}}{\Gamma \left( \frac{d+1}{2} \right)}+ \frac{\pi^{1/2} \alpha \omega_{12}^{2}}{2 \Gamma \left( \frac{d+2}{2} \right)} - \frac{(1-\alpha^{2}) \omega_{12}^{3}}{4 \Gamma \left( \frac{d+3}{2} \right)} \right] \phi(\omega_{1}) \phi (\omega_{2})=0.
\end{equation}
At this point, an approximated expression of the velocity distribution $\phi (\omega)$ will be introduced. To formulate it in a simple way, introduce a new velocity variable,
\begin{equation}
\label{4.28}
{\bm c} \equiv \frac{\bm \omega}{\omega_{0}}.
\end{equation}
Here, $\omega_{0}$ is the thermal velocity relative to the characteristic speed $\Delta$,
\begin{equation}
\label{4.29}
\omega_{0} \equiv \left( \frac{2T_{s}}{m} \right) ^{1/2}\, \frac{1}{\Delta}.
\end{equation}
The normalized to unity distribution of the ${\bm c}$ velocities is given by
\begin{equation}
\label{4.30}
\varphi(c) = \omega_{0}^{d} \phi (\omega),
\end{equation}
and its second moment is
\begin{equation}
\label{4.31}
\int d{\bm c}\, c^{2} \varphi (c) = \frac{d}{2}.
\end{equation}
Now, the function $\varphi (c)$ is expanded in Sonine polynomials as \cite{RydL77}
\begin{equation}
\label{4.32}
\varphi(c) = \varphi^{(0)} (c) \sum_{j=0}^{\infty} a_{j} S^{(j)} (c^{2}),
\end{equation}
where
\begin{equation}
\label{4.33}
\varphi^{(0)}= \pi^{-d/2} e^{- c^{2}}\,
\end{equation}
and the Sonine polynomials (closely related to the associated Laguerre polynomials) are defined by
\begin{equation}
\label{4.34}
S^{(j)} (x) = \sum_{r=0}^{j} \frac{\Gamma \left( j+d/2 \right)}{(j-r)! r! \Gamma \left( r+d/2 \right)}\, (-x)^{r}.
\end{equation}
The Sonine polynomials verify the orthogonality condition
\begin{equation}
\label{4.35}
\int d{\bm c}\, \varphi^{(0)}(c) S^{(j)}(c^{2}) S^{(j^{\prime})} (c^{2}) = \frac{\Gamma \left( j+d/2 \right)}{\Gamma \left( d/2 \right) j!}\, \delta_{j,j^{\prime}}.
\end{equation}
Normalization of $\varphi (c)$ and Eq.\ (\ref{4.31}) imply that $a_{0}=1$ and $a_{1}=0$, respectively. The next coefficient in the expansion (\ref{4.32}), $a_{2}$, is related with the forth moment of the scaled distribution,
\begin{equation}
\label{4.36}
a_{2} \equiv \frac{4 <c^{4}>}{d(d+2)}\, -1,
\end{equation}
with
\begin{equation}
\label{4.37}
<c^{4}> \equiv \int d{\bm c}\, c^{4} \varphi (c).
\end{equation}
In the following, the first Sonine approximation of the scaled distribution function,
\begin{equation}
\label{4.38}
\varphi(c)\approx \varphi^{(0)}(c) \left[ 1 + a_{2} S^{(2)} (c^{2}) \right],
\end{equation}
will be  considered. Note that because of the scaling of the probability density of the steady state, the coefficient $a_{2}$ does not depend on the velocity parameter  $\Delta$. Later on, it will be discussed how the parameter $a_{2}$ can be determined. For the moment, let us assume that it is $|a_{2}| \ll 1$, so that nonlinear in $a_{2}$ terms can be safely neglected, at least when considering low order velocity moments. This assumption must be checked {\em a posteriori} from the consistency of the results obtained and also by measuring $a_{2}$ by means of computer simulation methods. Substitution of Eq.\ (\ref{4.38}) into Eq.\ (\ref{4.27}) leads, in the aforementioned linear in $a_{2}$  approximation, to
\begin{equation}
\label{4.39}
\frac{1-\alpha^{2}}{2} \left ( 1+\frac{3 a_{2}}{16} \right) \omega_{0}^{2}-\left( \frac{\pi}{2} \right)^{1/2} \alpha \omega_{0} -1 + \frac{a_{2}}{16}=0.
\end{equation}
This equation provides a formal expression for $\omega_{0}$ and, therefore, for the steady temperature $T_{s}$. The expression involves the until now unknown parameter $a_{2}(\alpha)$. The positive root of the equation, keeping consistently only up to linear terms in $a_{2}$ reads
\begin{equation}
\label{4.40}
\omega_{0} \approx \frac{\pi^{1/2} \alpha}{2^{1/2} (1-\alpha^{2})} \left[ A(\alpha)+a_{2} (\alpha) B(\alpha) \right],
\end{equation}
with
\begin{equation}
\label{4.41}
A(\alpha) = 1 + \left[ \frac{4-(4-\pi)\alpha^{2}}{\pi \alpha^{2}} \right]^{1/2},
\end{equation}
\begin{equation}
\label{4.42}
B(\alpha)= - \frac{8(1-\alpha^{2})+ 3 \pi \alpha^{2}}{16 \alpha}\, \left\{ \frac{1}{\left[ 4-(4-\pi)\alpha^{2} \right]\pi } \right\}^{1/2}- \frac{3}{16}\, .
\end{equation}
If the term proportional to $a_{2}$ is neglected, i.e. $\varphi(c)$ is approximated by simply the Gaussian $\varphi^{(0)}(c)$, Eq.\ (\ref{4.40}) yields
\begin{equation}
\label{4.43}
T_{s} \approx T_{s}^{(G)} = \frac{m \alpha^{2} \pi}{4(1-\alpha^{2})^{2}} \left[ 1 + \sqrt{1+ \frac{4(1-\alpha^{2})}{\pi \alpha^{2}}} \right]^{2} \Delta^{2}.
\end{equation}
This expression agrees with the one reported in \cite{BRyS13}, except for the factor $m$, which is taken unity there.

To obtain an expression for the coefficient $a_{2}(\alpha)$, a procedure simular to the one employed to derive Eq.\ (\ref{4.40}) will be used. Nevertheless, this time  the Enskog equation will be considered from the beginning, for the sake of simplicity. For the steady state, Eq.\ (\ref{3.20}) reduces to
\begin{eqnarray}
\label{4.44}
\int d{\bm \omega}_{2} \int d \widehat{\bm \sigma}\, \left[ \theta \left( {\bm \omega}_{12} \cdot \widehat{\bm \sigma} -2 \right) \left( {\bm \omega}_{12} \cdot \widehat{\bm \sigma} -2 \right) \alpha^{-2} b_{\bm \sigma}^{-1} (1,2) \right. \nonumber \\
\left. - \theta \left( {\bm \omega}_{12} \cdot \widehat{\bm \sigma} \right)\left( {\bm \omega}_{12} \cdot \widehat{\bm \sigma} \right) \right] \phi (\omega_{1}) \phi (\omega_{2}) =0.
\end{eqnarray}
This equation is now multiplied by $\omega_{1}^{4}$ and integrated over ${\bm \omega}_{1}$ to get
\begin{equation}
\label{4.45}
\int d{\bm \omega}_{1} \int d{\bm \omega}_{2} \int d \widehat{\bm \sigma}\,  \phi (\omega_{1}) \phi (\omega_{2}) \theta ( -{\bm \omega}_{12} \cdot \widehat{\bm \sigma}) |{\bm \omega}_{12} \cdot \widehat{\bm \sigma}| \left[ b_{\bm \sigma}(1,2) -1 \right] (\omega_{1}^{4}+ \omega_{2}^{4})=0.
\end{equation}
Next, the angular integral is carried out to get an expression of the form
\begin{equation}
\label{4.46}
\int d \widehat{\bm \sigma}\, \theta ( -{\bm \omega}_{12} \cdot \widehat{\bm \sigma}) |{\bm \omega}_{12} \cdot \widehat{\bm \sigma}| \left[ b_{\bm \sigma}(1,2) -1 \right] (\omega_{1}^{4}+ \omega_{2}^{4}) = K({\bm \omega}_{12}, {\bm W}) ,
\end{equation}
where ${\bm W} \equiv ({\bm \omega}_{1}+ {\bm \omega}_{2})/2$. In order to introduce the first Sonine approximation, it is again convenient to formulate the problem in the ${\bm c}$-velocity scale defined by Eq.\ (\ref{4.28}). Then, Eq. (\ref{4.45}) reads
\begin{equation}
\label{4.47}
\int d{\bm c}_{1} \int d{\bm c}_{2}\, \varphi (c_{1}) \varphi(c_{2}) K_{1}({\bm c}_{12} ,{\bm G}) =0,
\end{equation}
with ${\bm G} \equiv {\bm W}/\omega_{0}$, $K_{1}({\bm c}_{12} ,{\bm G}) \equiv K({\bm \omega}_{12}, {\bm W})$, and $\varphi (c)$ defined by Eq.\ (\ref{4.30}). Note that the expression of $K_{1}$ involves powers of $\omega_{0}$. To evaluate the left hand side in the above equation, the first Sonine approximation as defined by Eq. (\ref{4.38}) is used. Moreover, consistently with the previous calculations, terms proportional to $a_{2}^{2}$ are neglected, i.e. Eq. (\ref{4.47}) is approximated by
\begin{equation}
\label{4.48}
\int d{\bm c}_{1} \int d{\bm c}_{2}\, \varphi^{(0)} (c_{1}) \varphi^{(0)}(c_{2}) \left\{ 1 + a_{2}\left[ S^{(2)} (c_{1}^{2}) + S^{(2)} (c_{2}^{2}) \right] \right\} K_{1}({\bm c}_{12} ,{\bm G}) =0.
\end{equation}
Now the computational problem has been reduced to evaluate Gaussian velocity integrals. This is a lengthly but easy calculation, and Eq. (\ref{4.48}) becomes an equation for $a_{2}$ and $\omega_{0}$. But an expression for $\omega_{0}$ valid to linear order in $a_{2}$ was derived above, Eq. (\ref{4.40}). Then, this expression is substituted into the result of evaluating the left hand side of Eq.\ (\ref{4.48}) and, once again, only terms of zeroth and first order in $a_{2}$ are kept. In this way, a linear equation for the latter is obtained, leading to an expression of it as a function of the coefficient of normal restitution $\alpha$.

All the calculation just described are easily done by using any of the available software for symbolic calculation. Nevertheless, the final result is rather long and not very illuminating and, therefore, it will be not reproduced here. Instead, the first few terms in the expansion of $a_{2}$ in powers of $1-\alpha^{2}$ is given below. As dicussed in the next section, this analytical expression is quite accurate for not too strong inelasticity. Keeping up to order $(1- \alpha^{2})^3$, it is
\begin{equation}
\label{4.49}
a_2 \approx p_1(1-\alpha^2)+p_2(1-\alpha^2)^2+p_3(1-\alpha^2)^3\, ,
\end{equation}
with the coefficients $p_{1}$, $p_{2}$, and $p_{3}$ given by
\begin{equation}
\label{4.50}
p_1=-\frac{1}{4(d-1)}\, ,
\end{equation}
\begin{equation}
\label{4.51}
p_2=\frac{-24+d(24-7\pi)+13\pi}{32(d-1)^2\pi}\, ,
\end{equation}
\begin{equation}
\label{4.53}
p_3=\frac{-384+164\pi-49\pi^2+2d(384-130\pi+\pi^2)+d^2(-384+96\pi+11\pi^2)}{256(d-1)^3\pi^2}\, ,
\end{equation}
respectively.

\section{Simulation results}
\label{s5}
To test the accuracy of the theoretical predictions derived in the previous section, the DSMC method \cite{Bi94,Ga00} has been used to generate numerical solutions of the Boltzmann equation in the two-dimensional case. The only modification needed to adapt the simulation of elastic hard disks to the dynamics of the model is the collision rule, Eqs.\ (\ref{2.1}) and (\ref{2.2}). Attention here will be restricted to the properties of the steady homogeneous state discussed in Sec.\ \ref{s4}. Therefore, when implementing the DSMC method, the position of the particles is not relevant and it is sufficient to consider only one spatial cell. Moreover, the difference between the Enskog and the Boltzmann equation lies in the equilibrium pair correlation function at contact, $g_{e}(\sigma;n)$, which is a constant for the steady state. That means that solutions of the homogeneous Boltzmann equation, and therefore obtained with $g_{e}=1$, are trivially translated into solutions of the homogeneous Enskog equation.

In the simulation reported below, a system of typically $N=1000$ particles has been employed. It must be kept in mind that the number of particles used in the DSMC method only has an statistical meaning, and does not affect the validity of the low-density limit, which is inherent to the method itself \cite{Bi94}. To improve the statistics, the results have been averaged over independent realizations or replicas. Its number has changed between $10$, for the moments of the distribution, and  $5000$, for the velocity distributions themselves. The initial velocity distribution of the disks was always a Gaussian with an initial temperature $T(0)=T_{0}$. In all cases, it was observed that the system reached a steady state, characterized by constant properties that are independent from the initial state. The transient time for the relaxation to the steady state is of the order of a few collisions per particle.

\begin{figure}
\includegraphics[scale=0.4,angle=0]{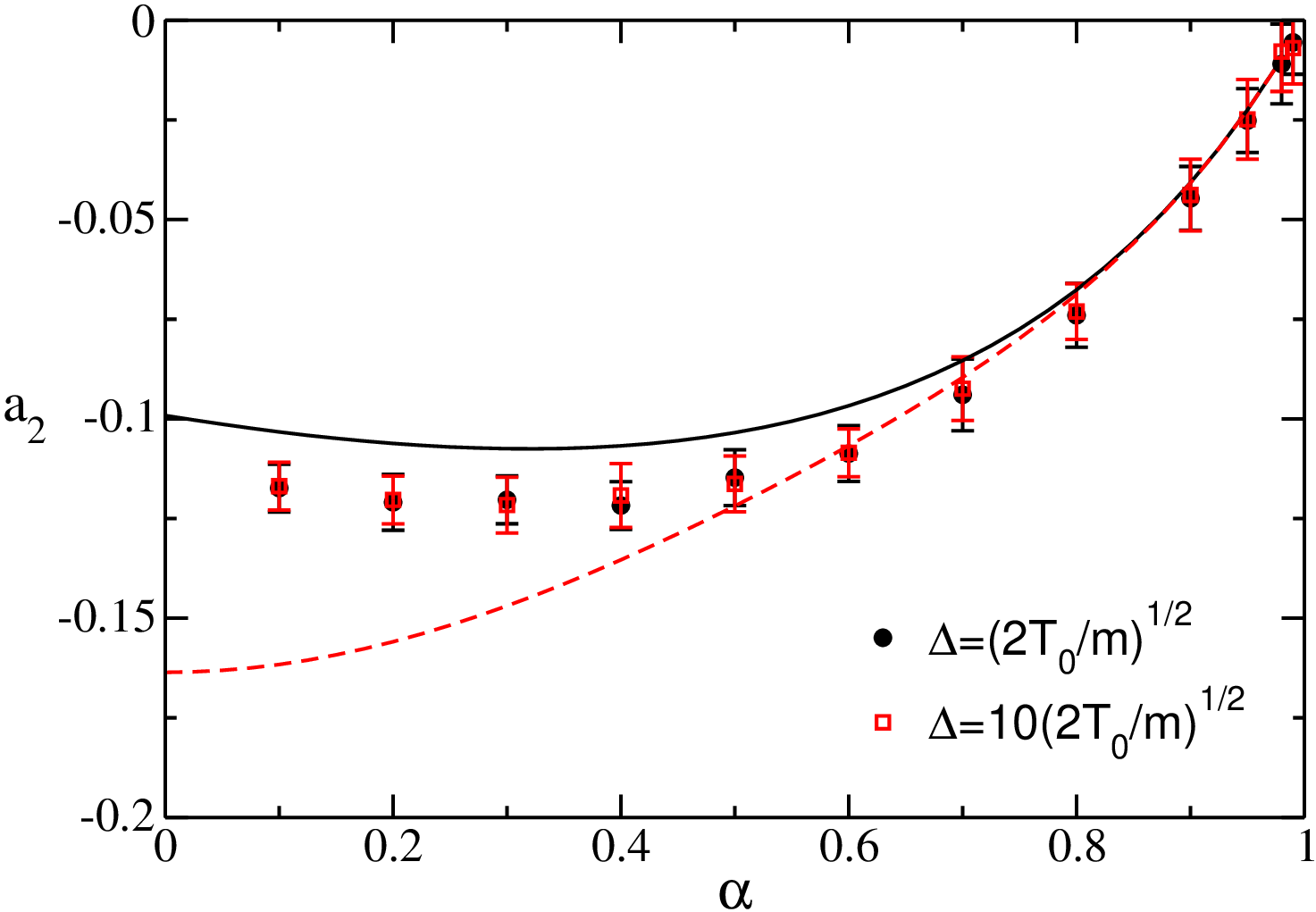}
\caption{(Color online) Dimensionless first Sonine coefficient $a_{2}$, as a function of the coefficient of normal restitution $\alpha$, for a system of repulsive inelastic hard disks. The symbols are simulation results obtained by the DSMC method, for two different values of the velocity $\Delta$, as indicated in the inset, the solid line is the theoretical prediction using the complete $\alpha$ dependence, and the dashed line is the result obtained for small inelasticity, as given in Eq. (\protect{\ref{4.49}}).  \label{fig1}}
\end{figure}

The results for the Sonine coefficient $a_{2}(\alpha)$ are given in Fig.\ \ref{fig1}. Data for two different values of the velocity $\Delta$ have been given, namely $ \Delta= \left( 2 T_{0}/m \right)^{1/2} $ and $\Delta = 10 \left( 2 T_{0}/m \right)^{1/2} $, respectively. Both series of data are indistinguishable within the statistical uncertainties, indicating that $a_{2}$ is independent from $\Delta$, as predicted. Also plotted (solid line) in the figure is the theoretical prediction for $a_{2} (\alpha)$ obtained in the first Sonine approximation, whose analytical expression has not been reported here. A fairly good agreement is observed over all the range of values of the coefficient of normal restitution $\alpha$. Finally, the small inelasticity approximation, given by Eqs. (\ref{4.49})-(\ref{4.53}) is included (dashed line). It is seen that for not too strong inelasticity, $ \alpha \agt 0.5$, this expression provides a fairly good approximation.

\begin{figure}
\includegraphics[scale=0.4,angle=0]{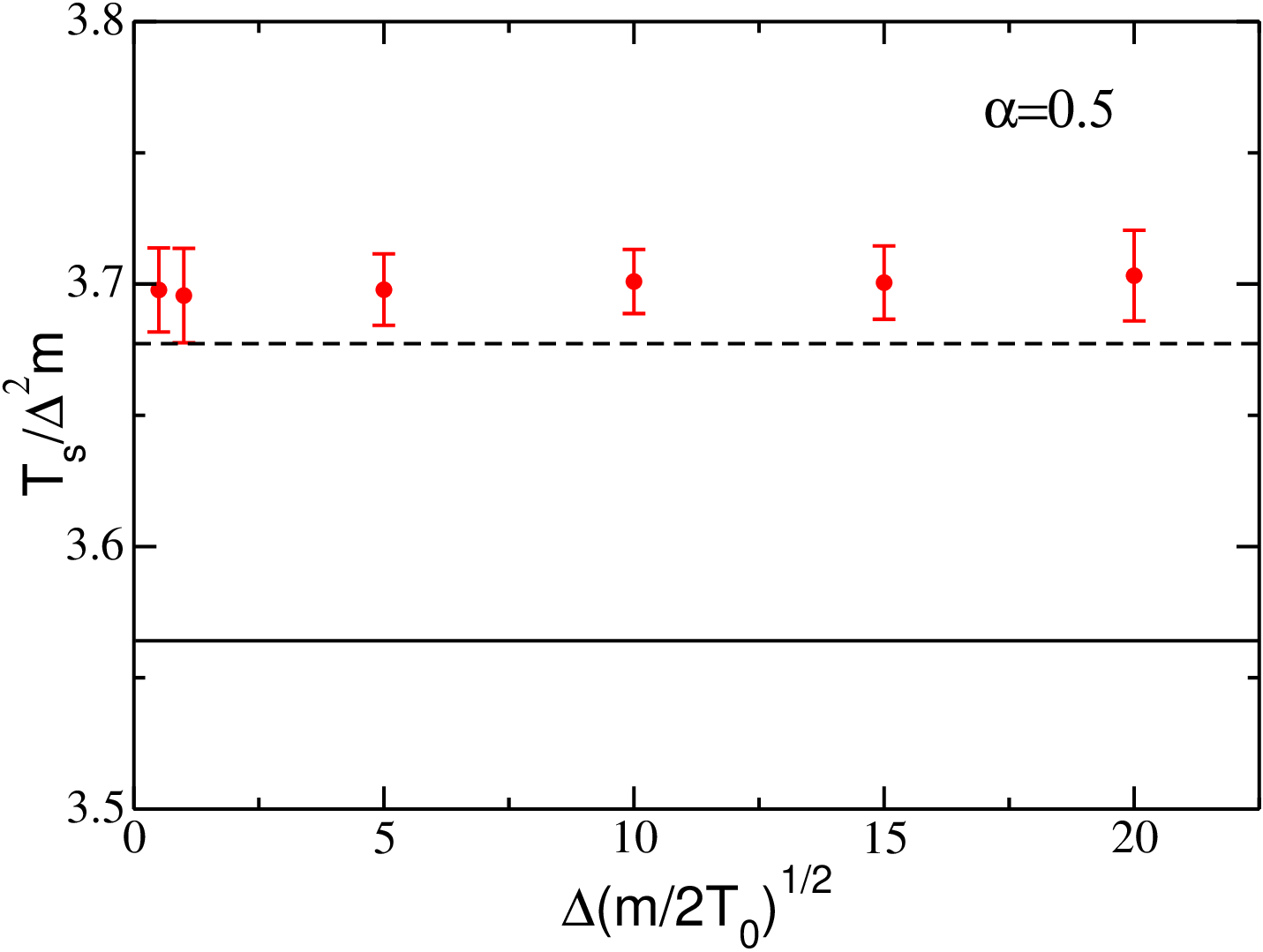}
\caption{(Color online) Normalized dimensionless steady temperature for a system of repulsive inelastic hard disks with $\alpha=0.5$.  The (red) symbols are numerical results obtained with the DSMC method, the solid line the theoretical prediction obtained with a Gaussian velocity distribution, Eq.\ (\protect{\ref{4.43}}), and the dashed  line is the result obtained using the first Sonine approximation for the velocity distribution, Eq.\ (\protect{\ref{4.40}}). \label{fig2}}
\end{figure}

\begin{figure}
\includegraphics[scale=0.4,angle=0]{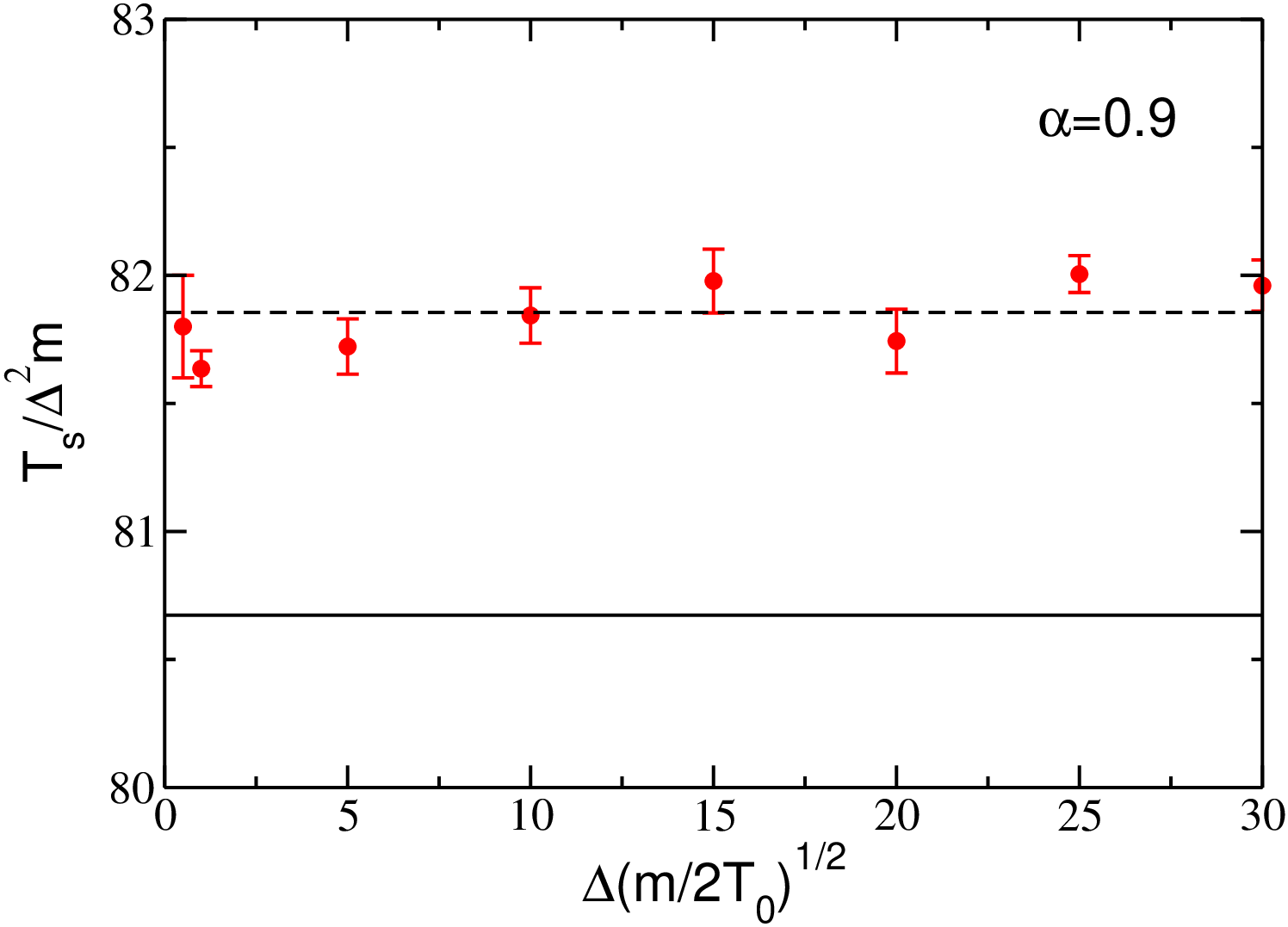}
\caption{(Color online) The same as in Fig.  \protect{\ref{fig2}}, but now for a system of repulsive inelastic hard disks with $\alpha=0.9$. \label{fig3}}
\end{figure}

\begin{figure}
\includegraphics[scale=0.4,angle=0]{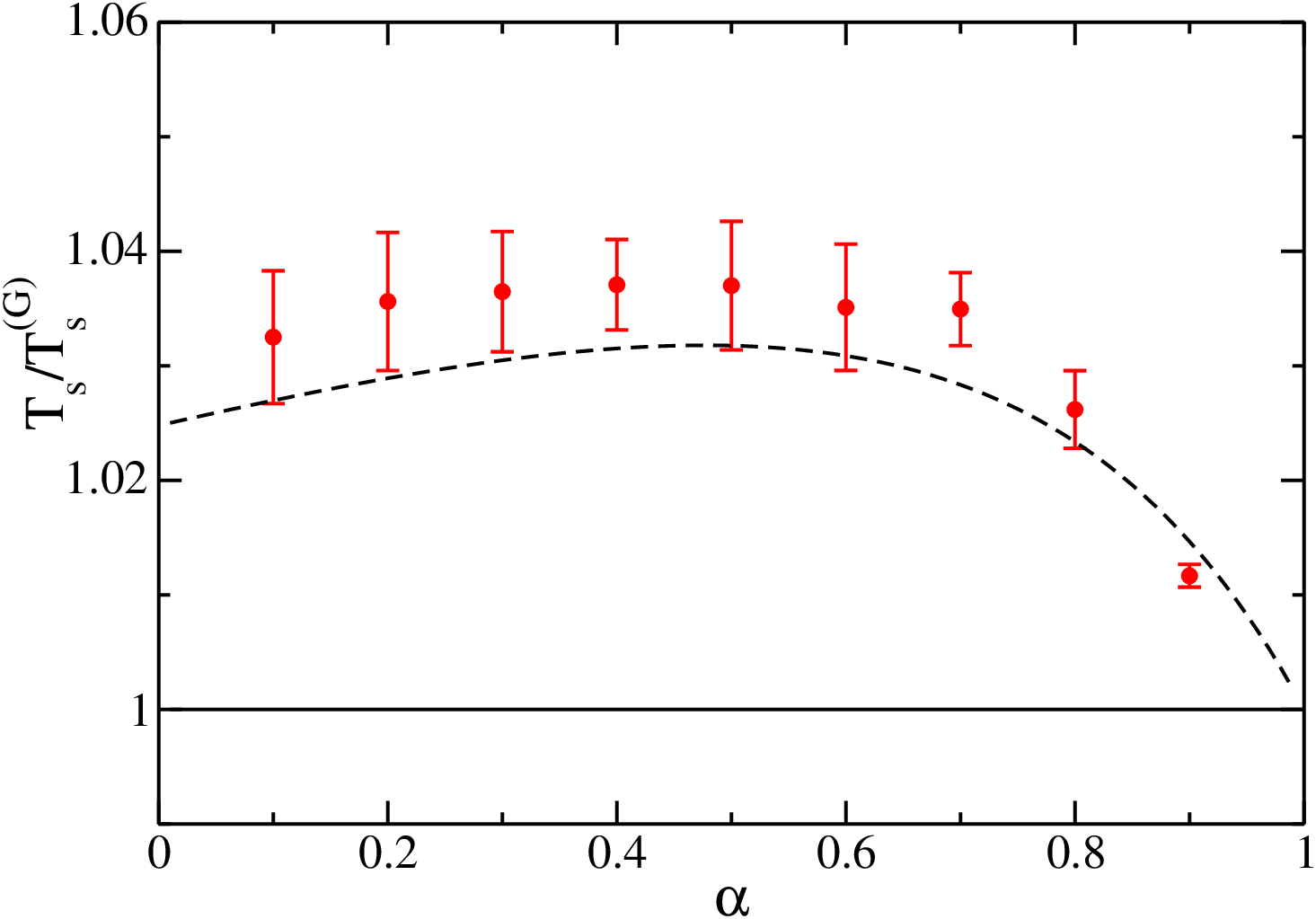}
\caption{(Color online) The steady temperature $T_{s}$, normalized with the theoretical prediction using a Gaussian velocity distribution, $T_{s}^{(G)}$, as a function of the coefficient of normal restitution $\alpha$. The dashed curve is the theoretical prediction in the first Sonine approximation, while the (red) symbols are simulation results. The horizontal line corresponding to the Gaussian approximation is just a guide for the eye. \label{fig3a}}
\end{figure}

Concerning the influence of non-Gaussian effects on the steady temperature, Figs. \ref{fig2} and \ref{fig3} show the steady temperature for systems with $\alpha=0.5$ and $\alpha=0.9$, respectively. In both cases it is seen that the steady temperature does not depend on the velocity parameter $\Delta$, as predicted. The solid lines are the values of the steady temperature obtained by means of a Gaussian velocity distribution, i.e. those given by Eq.\ (\ref{4.43}), while the values indicated by the dashed lines are obtained by considering the first Sonine approximation, Eq.\ (\ref{4.40}).
In the latter case, the theoretical prediction for $a_{2}$ has been employed. The conclusion is that the influence of the non-Gaussianity of the distribution function on the steady temperature, although it is small, is clearly distinguishable, and it is fairly well described by the first Sonine approximation considered in this paper. This is summarized in Fig.\ \ref{fig3a}, where the steady temperature $T_{s}$ divided by the theoretical prediction obtained in the Gaussian approximation $T_{s}^{(G)}$, Eq.\ (\ref{4.43}), is plotted as a function of the coefficient of normal restitution $\alpha$. the value of the parameter $\Delta$ is not specified, since the temperature ratio is known to be independent from it.

\begin{figure}
\includegraphics[scale=0.4,angle=0]{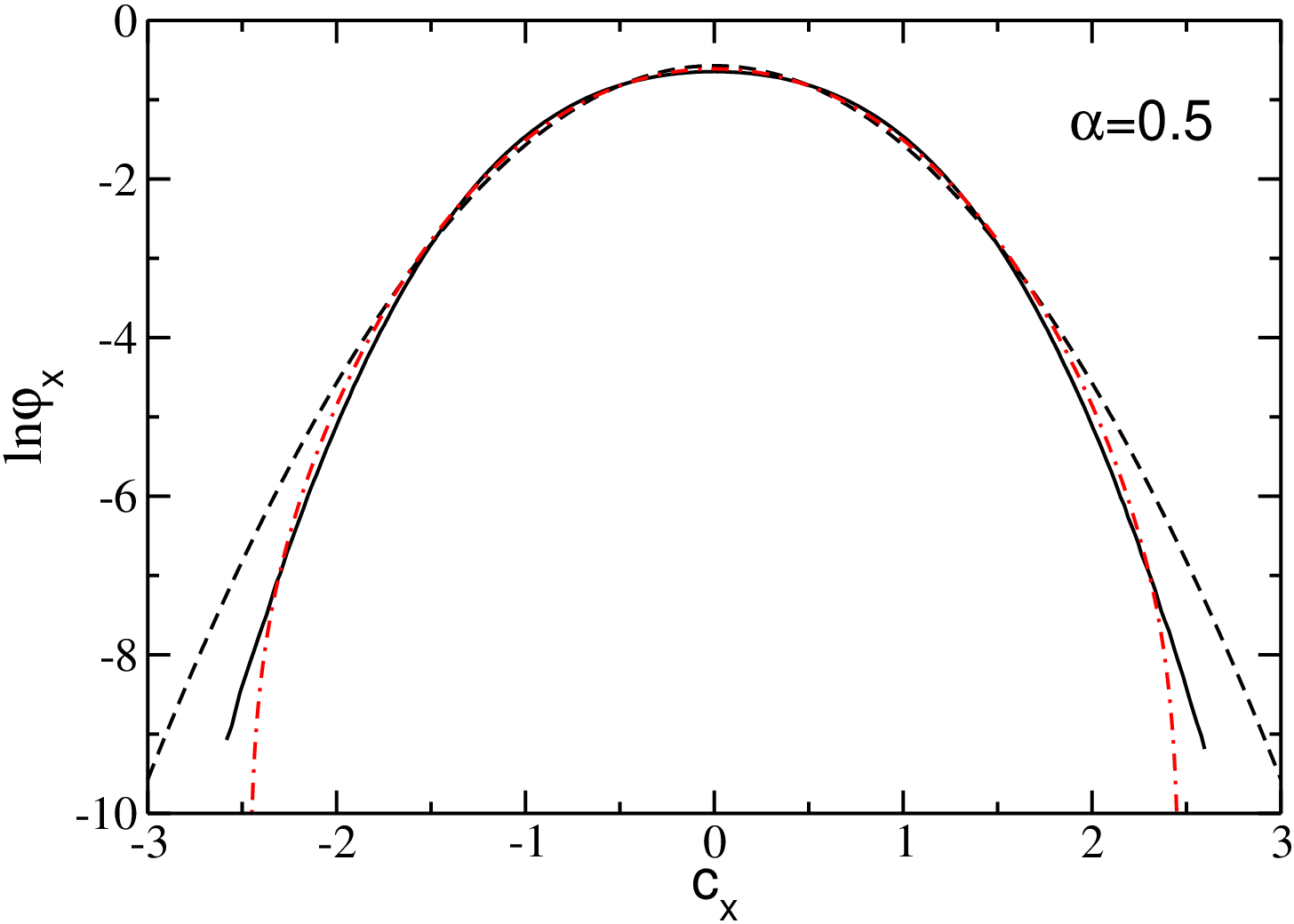}
\caption{(Color online) Marginal velocity distribution $\varphi_{x}(c_{x})$ of the steady state of a granular fluid of repulsive hard disks with $\alpha=0.5 $. The (black) solid line represents the simulation results, the (black) dashed line is the Maxwell distribution, and the (red) dot-dashed line is the result obtained in the first Sonine approximation, as discussed in the main text.  \label{fig4}}
\end{figure}

\begin{figure}
\includegraphics[scale=0.4,angle=0]{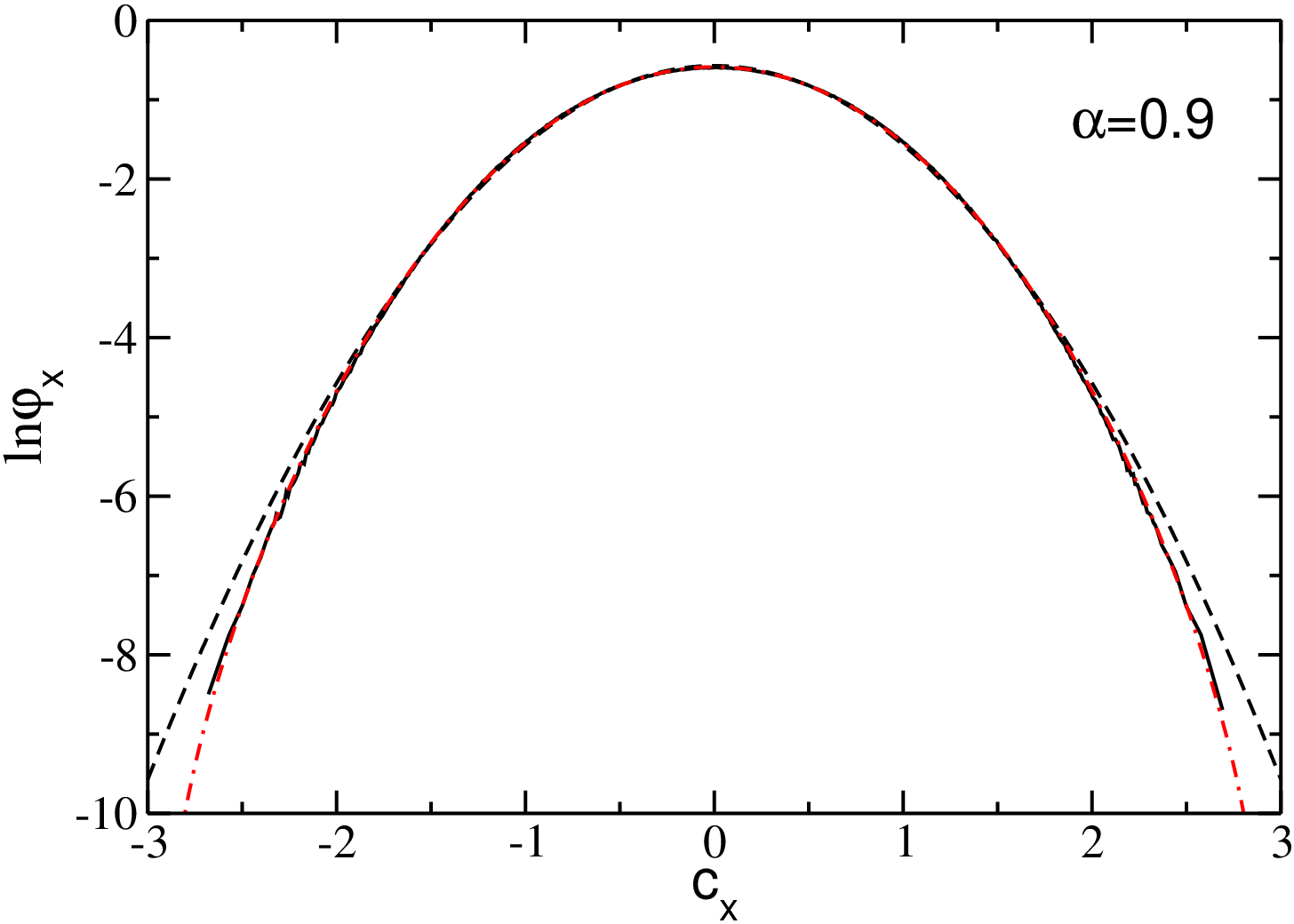}
\caption{(Color online) The same as in Fig.  \protect{\ref{fig4}}, but  for a system of repulsive inelastic hard disks with $\alpha=0.9$. \label{fig5}}
\end{figure}

To test whether the first Sonine approximation also provides a good estimation for the one-particle distribution function of the steady state, the normalized marginal distribution function
\begin{equation}
\label{5.1}
\varphi_{x}(c_{x}) =  \int_{-\infty}^{+\infty} dc_{y}\, \varphi (c),
\end{equation}
has been considered. Figures \ref{fig4} and \ref{fig5} show $\varphi_{x} (c_{x})$ for $\alpha =0.5$ and $\alpha=0.9$, respectively. Note that a semi-logarithmic representation is being used. Also plotted is the Maxwellian, $\varphi^{(0)}(c_{x}) = (\pi)^{-1/2} \exp(-c_{x}^{2})$, and the first Sonine approximation,
\begin{equation}
\label{5.2}
\varphi_{x}(c_{x})\approx \varphi^{(0)} (c_{x}) \left[ 1 + \frac{a_{2}}{2} \left( c_{x}^{4}-3 c_{x}^{2} + \frac{3}{4} \right) \right].
\end{equation}
The value used for $a_{2}$ was the one obtained by the method described in Sec.\ \ref{s4}. Two main conclusions follow from these figures. Firstly, that the distortion of the velocity distribution from the Maxwellian is clearly identifiable, even for weak dissipation and, secondly, that the distorsion is quite accurate described by the first Sonine approximation, even for rather strong dissipation.

\section{Discussion}
\label{s6}
 In this paper, the statistical mechanics and kinetic theory for a model recently introduced by Brito {\em et al.} \cite{BRyS13} to describe the dynamics of a confined granular gas of hard spheres or disks, have been formulated. Expressions for the dynamics of phase functions as well as for the distribution function of the system have been derived. The model involves a characteristic constant speed $\Delta$ that is added to the relative velocity along the normal direction in each collision. It has been shown that this speed can be eliminated from the equation describing the dynamics by means of a change in the time scale.

The pseudo-Lioville equation governing the time evolution of the distribution function of the system admits a stationary solution in which the energy dissipated in collisions is balanced with the energy injected through the impulsive velocity $\Delta$. As a consequence of the scaling mentioned above, the dependence  of the moments of the steady distribution on $\Delta$ is easily identified. This includes as a particular case the temperature of the system.

To derive explicit expressions for the temperature, and also for the velocity distribution of the steady state, kinetic theory was used, in the context of the Enskog approximation. Moreover, an expansion in Sonine polynomials, truncated to the lowest non-trivial order, was employed. In this way, the non-equilibrium effects implied by the deviations from a Gaussian of the distribution function were investigated. A point to be emphasized is that the Enskog theory also provides relevant information about the velocity  correlations generated by the collisions in non-equilibrium states \cite{Lu99}. These correlations in turn also lead to additional contributions to the spatial correlations. The expression of the pair correlations of two particles at contact is given by Eq.\ (\ref{3.24}). More specifically, particularization of that equation to the steady state gives (see Appendix \ref{ap1})
\begin{equation}
\label{6.1}
g_{s}({\bm r}_{1},{\bm r}_{2}) \delta (r_{12}- \sigma)= \left[ \frac{1+\alpha}{2 \alpha} - h_{s}(\alpha) \right] g_{e}(\sigma;n) \delta (r_{12} -\sigma),
\end{equation}
\begin{equation}
\label{6.2}
h_{s}(\alpha) = \alpha^{-1} \int d{\bm c}_{1} \int d{\bm c}_{2}\, \frac{1}{2+ \alpha \omega_{0} |{\bm c}_{12} \cdot \widehat{\bm q}_{12}|}\, \varphi (c_{1}) \varphi (c_{2})\, .
\end{equation}

The theoretical predictions have been compared with numerical solutions obtained by the direct simulation Monte Carlo method, and a very good agreement has been found. In particular, it has been shown that the first Sonine approximation provides a fairly good description of the one-particle velocity distribution of the system, even for rather strong inelasticity. In this context, it is worth mentioning that the arguments leading to the existence of a high energy tail in the velocity distribution of a granular gas in the homogeneous cooling state \cite{EyP97,BCyR99}, can not be applied to the steady state displayed by the present model.

To put the present work in a proper context, it is important to notice that attention here has been restricted to homogeneous states. Consistently, the stability of the steady state has not been investigated. In particular, no spatial fluctuations have been allowed in the DSMC simulations. Let us mention that by means of a general hydrodynamic argument it was established in ref. \cite{BRyS13} that the steady state should always be stable. As a consequence, it was concluded that the model is not able to reproduce the solid-liquid transition observed in quasi-two-dimensional systems \cite{RPGRSCyM11,OyU05}. Nevertheless, we believe that this issue deserves some additional attention, trying to clarify what happens in the present model  with the solid-fluid transition exhibited by elastic hard spheres.

\section{Acknowledgements}

This research was supported by the Ministerio de Educaci\'{o}n y Ciencia (Spain) through Grant No. FIS2011-24460 (partially financed by FEDER funds).

\appendix*

\section{The pair correlation function at contact}
\label{ap1}

Decompose $f_{2}({\bm x}_{1},{\bm x}_{2},t)$ in the form
\begin{equation}
\label{ap1.1}
f_{2}({\bm x}_{1},{\bm x}_{2},t)= f_{2}^{(+)}({\bm x}_{1},{\bm x}_{2},t) +f_{2}^{(-)}({\bm x}_{1},{\bm x}_{2},t)
\end{equation}
with
\begin{equation}
\label{ap1.2}
f_{2}^{+}({\bm x}_{1},{\bm x}_{2},t) \equiv \theta ({\bm v}_{12} \cdot \widehat{\bm r}_{12}) f_{2}({\bm x}_{1},{\bm x}_{2},t),
\end{equation}
\begin{equation}
\label{ap1.3}
f_{2}^{(-)}({\bm x}_{1},{\bm x}_{2},t) \equiv  \theta (-{\bm v}_{12} \cdot \widehat{\bm r}_{12}) f_{2}({\bm x}_{1},{\bm x}_{2},t).
\end{equation}
Use of Eq.\ (\ref{3.18}) into Eq.\ (\ref{3.23}) yields
\begin{equation}
\label{ap1.4}
f_{2}^{(+)}({\bm x}_{1},{\bm x}_{2},t) | {\bm v}_{12} \cdot \widehat{\bm r}_{12} | \delta (r_{12}-\sigma ) = \alpha^{-2} | {\bm v}_{12} \cdot \widehat{\bm r}_{12} - 2 \Delta | b_{\bm r}^{-1} (1,2) f_{2}^{(-)}( {\bm x}_{1},{\bm x}_{2},t) \delta (r_{12}-\sigma).
\end{equation}
This equation can be understood as expressing the boundary condition for the free motion of the hard spheres or disks as a consequence of the collisions between them \cite{Lu99}. Now the RET is introduced. Substitution of Eq.\ (\ref{3.19}) on the right hand side of Eq.\ (\ref{ap1.4}) gives
\begin{eqnarray}
\label{ap1.5}
f_{2}^{(+)}({\bm x}_{1},{\bm x}_{2},t) \delta (r_{12}- \sigma) &=& \alpha^{-2} \frac{| {\bm v}_{12} \cdot \widehat{\bm r}_{12} - 2 \Delta |}{| {\bm v}_{12} \cdot \widehat{\bm r}_{12} |} b_{\bm r}^{-1} (1,2) f_{1} ({\bm x}_{1},t)f_{1}({\bm x}_{2},t) \nonumber \\
& & \times g_{E}({\bm r}_{1},{\bm r}_{2},t) \theta (- {\bm v}_{12} \cdot \widehat{\bm r}_{12}) \delta (r_{12}- \sigma).
\end{eqnarray}
Consequently, in the RET it is
\begin{eqnarray}
\label{ap1.6}
f_{2}({\bm x}_{1},{\bm x}_{2},t) \delta (r_{12}-\sigma) & = & \left\{ 1 + \left[ \alpha^{-2} \frac{| {\bm v}_{12} \cdot \widehat{\bm r}_{12} - 2 \Delta |}{| {\bm v}_{12} \cdot \widehat{\bm r}_{12} |} \theta \left( {\bm v}_{12} \cdot \widehat{\bm r}_{12}  - 2 \Delta \right) b_{\bm r}^{-1}(1,2) \right. \right.
\nonumber \\
& & \left. \left. - \theta ( {\bm v}_{12} \cdot \widehat{\bm r}_{12} ) \right] \right\}
f_{1} ({\bm x}_{1},t)f_{1}({\bm x}_{2},t) g_{E}({\bm r}_{1},{\bm r}_{2},t)  \delta (r_{12}- \sigma).
\end{eqnarray}

The pair correlation function $g({\bm r}_{1},{\bm r}_{2},t)$ is defined by
\begin{equation}
\label{ap1.7}
n({\bm r}_{1},t) n ({\bm r}_{2},t) g({\bm r}_{1},{\bm r}_{2},t)= \int d{\bm v}_{1} \int d{\bm v}_{2}\ f_{2} ({\bm x}_{1},{\bm x}_{2},t).
\end{equation}
Then, integration over the velocities  of Eq.\ (\ref{ap1.6}) leads directly to Eq. (\ref{3.24}).

Suppose now that the state being considered is isotropic in velocity, so that $f_{1}({\bm x},t)$ only depends on velocity through its modulus $|{\bm v}|$. Then, Eq.\ (\ref{3.24}) can be transformed into
\begin{eqnarray}
\label{ap1.8}
g({\bm r}_{1},{\bm r}_{2},t) \delta ( r_{12}- \sigma)& = & g_{E}({\bm r}_{1},{\bm r}_{2},t) \delta ( r_{12} -\sigma)
+ \frac{1}{n_{1}({\bm r}_{1},t) n({\bm r}_{2},t)} \nonumber \\
 && \times \int d{\bm v}_{1} \int d{\bm v}_{2} \left( \frac{| {\bm v}_{12} \cdot \widehat{\bm r}_{12} |}{ 2 \Delta + \alpha | {\bm v}_{12} \cdot \widehat{\bm r}_{12}|}\, - 1 \right)\theta ({\bm v}_{12} \cdot \widehat{\bm r}_{12} ) \nonumber \\
&& \times f({\bm x}_{1},t) f({\bm x}_{2},t) g_{E}({\bm r}_{1},{\bm r}_{2},t) \delta (r_{12}-\sigma),
\end{eqnarray}
or, equivalently,
\begin{equation}
\label{ap1.9}
g({\bm r}_{1},{\bm r}_{2},t) \delta ( r_{12}- \sigma) = \left[ \frac{1+\alpha}{ 2 \alpha} - h (\alpha, \Delta, t) \right] g_{E}({\bm r}_{1},{\bm r}_{2},t) \delta (r_{12}-\sigma),
\end{equation}
with
\begin{equation}
\label{ap1.10}
h(\alpha, \Delta,t) = \frac{\alpha^{-1} \Delta}{n({\bm r}_{1},t) n({\bm r}_{2},t)} \int d{\bm v}_{1} \int d{\bm v}_{2}\, \frac{1}{2 \Delta+ \alpha | {\bm v}_{12} \cdot \widehat{\bm r}_{12}| }\, f_{1} ({\bm x}_{1},t) f_{1} ({\bm x}_{2},t).
\end{equation}
Equation (\ref{6.1}) is the particularization of this expression for the steady state.

\end{document}